\documentclass[aip,jcp,reprint]{revtex4-1}

\usepackage{color}

\usepackage{gensymb}
\usepackage{graphicx}
\usepackage{dcolumn}
\usepackage{bm}
\usepackage{amsmath}
\usepackage{amsthm}
\usepackage{verbatim}
\usepackage{mciteplus}
\usepackage[welsh,english]{babel}

\begin{document}

\title{The water supercooled regime as described by four common water models}

\author{David C. Malaspina}
\affiliation{Department of Biomedical Engineering, Northwestern University, 2145 Sheridan Road, Evanston, IL 60208}

\author{Aleida J. Berm\'udez di Lorenzo}
\affiliation{Facultad de Matem\'atica, Astronom\'{\i}a y F\'{\i}sica, Universidad Nacional de C\'ordoba, X5000HUA C\'ordoba, Argentina}

\author{Rodolfo G. Pereyra}
\affiliation{Facultad de Matem\'atica, Astronom\'{\i}a y F\'{\i}sica, Universidad Nacional de C\'ordoba, X5000HUA C\'ordoba, Argentina}
\affiliation{IFEG-CONICET, X5016LAE C\'ordoba, Argentina}

\author{Igal Szleifer}
\affiliation{Department of Biomedical Engineering, Northwestern University, 2145 Sheridan Road, Evanston, IL 60208}
\affiliation{Department of Chemistry, Northwestern University, 2145 Sheridan Road, Evanston, IL 60208}
\affiliation{Chemistry of Life Processes Institute, Northwestern University, 2145 Sheridan Road, Evanston, IL 60208}

\author{Marcelo A. Carignano}
\email{mcarignano@qf.org.qa}
\affiliation{Qatar Environment and Energy Research Institute, P.O. Box 5825, Doha, Qatar}
\affiliation{Department of Biomedical Engineering, Northwestern University, 2145 Sheridan Road, Evanston, IL 60208}

%\date{\today}

\begin{abstract}
The temperature scale of simple water models in general does not coincide with the natural one. Therefore, in order to make a meaningful evaluation of different water models a temperature rescaling is necessary. In this paper we introduce a rescaling using the melting temperature and the temperature corresponding to the maximum of the heat capacity to evaluate four common water models (TIP4P-Ew, TIP4P-2005, TIP5P-Ew and Six-Sites) in the supercooled regime. Although all the models show the same general qualitative behavior, the TIP5P-Ew appears as the best representation of the supercooled regime when the rescaled temperature is used. We also analyze, using thermodynamic arguments, the critical nucleus size for ice growth. Finally, we speculate on the possible reasons why atomistic models do not usually crystalize while the coarse grained mW model do crystallize. 
\end{abstract}

\maketitle

\newpage

\section{Introduction}

At atmospheric pressure, the solid-liquid transition temperature of water is $T_m=273.15$ K. Nevertheless, pure bulk water can be supercooled down to a temperature close to 235 K.\cite{Debenedetti:2003aa} Below that temperature, water crystallizes spontaneously to form hexagonal ice. While the melting temperature is well defined by the thermodynamic equilibrium of two phases, the lower limit of the supercooled regime is more difficult to define. Experiments with supercooled water show that the isobaric heat capacity, isothermal compressibility and thermal expansion coefficient appear to diverge as the temperature decreases below 240 K.\cite{ANGELL:1982aa,Tombari:1999aa,SPEEDY:1976aa,HARE:1986aa} Atomistic simulations of supercooled water present a different picture, although not necessarily in contradiction with the experiments. For example, the isobaric heat capacity calculated from simulations displays a maximum in the supercooled regime,\cite{Kumar:2007aa,Holzmann:2007aa,Pi:2009aa,Abascal:2010aa,Paula-Longinotti:2011aa} and no spontaneous crystallization is usually observed; except for one notable paper by Matsumoto et al.\cite{Matsumoto:2002aa} and the subsurface nucleation observed by Vrbka and Jungwirth.\cite{Vrbka:2006aa} The mW model, which is a coarse grained model for water, behaves in a different way than atomistic models. Spontaneous crystallization is observed in sufficiently long simulations using the mW model.\cite{Moore:2010aa,Moore:2011aa} 

Atomistic models are, in general, developed to reproduce experimental properties at certain thermodynamic conditions, usually ambient pressure and temperature. Nevertheless, it is common to explore the prediction of a model outside its initial target zone.\cite{Vega:2011aa} The case of supercooled water has attracted considerable attention of the simulation community, and several models were adapted to describe this regime. Yet, the proper capturing of the melting temperature $T_m$ by a simple model is difficult, and is customary in order to compare with experimental results, to use $T_m$ as a reference temperature and express the results in terms of the degree of supercooling, i.e. $T_m-T$.\cite{Shevchuk:2012ab} Nevertheless, given that the supercooled regime span over a wide temperature range, it is interesting to explore the prediction of different models using a rescaled variable based in the two temperature limits of the supercooled regime. In this way, any mistake that could be affecting the intrinsic energy scale of the models will be reduced or eventually removed and the comparison between different models and experiments becomes more meaningful.

The apparent divergence of the response functions has been interpreted in terms of several theoretical scenarios, which include a retracing spinodal of superheated water\cite{SPEEDY:1982aa}, a singularity free scenario\cite{Sastry:1996aa} and a first order liquid-liquid transition implying a second critical point in the metastable region\cite{POOLE:1992aa}. The proposed second critical point has gained significant support from recent simulation and experimental works. Indeed, extensive Monte Carlo simulations have shown the existence of this second critical point for ST2 water\cite{Liu:2012ac} and several experiments strongly suggest a phase transition between a high density liquid (HDL) and low density liquid (LDL).\cite{Liu:2007ab,Mallamace:2008aa,Zhang:2011aa} In spite of these results, the issue still remain controversial and a different interpretation has been proposed by Limmer and Chandler\cite{Limmer:2011aa}, who argue that the double basin observed in the  Monte Carlo simulations are a reflection of a liquid-crystal transition and not a liquid-liquid one. Within the second critical point scenario, the coexistence line between these two liquid states, which are metastable with respect to the crystalline phase, is called the Widom line and it is expressed in terms of pressure and temperature as $T_W(P)$. Then, the Widom line can be defined as the locus of the maxima of the heat capacity in the P-T plane and ends at the second critical point, which should be  in the vicinity of the 1 atm and 235 K in order to explain the apparent divergencies in the thermodynamic response functions mentioned above. Even if we assume that this scenario is correct, it does not univocally define the temperature of spontaneous nucleation, nor does it explain why a glassy state is not observed immediately below $T_W$. However, the second critical point scenario give us a way to define a low temperature reference point in order to compare experimental and simulation results using a proper rescaling. Yet, experiments in the deep supercooled regime are very difficult and therefore different measurements yield different temperatures for physical properties that reflect the HDL-LDL phase transition. For example, the work of  Mallamace et al.\cite{Mallamace:2008aa} shows that in water confined within $i$) micelle-templated mesoporous silica and $ii$) the hydration layer of lysozyme, the (negative) thermal expansion coefficient has a maximum at 238 K. Also, using NMR measurements of the proton chemical shift ($\delta$) they found a maximum in $-T(\partial \ln \delta/\partial T)_P$ at approximately the same temperature. This later quantity behaves similarly to $C_P$\cite{Mallamace:2008aa} and therefore is another indication of a phase transition between a high density and low density liquid. Maruyama et al.\cite{Maruyama:2004aa}, measured the maximum of the heat capacity at 227 K in water confined within silica gel pores. On the other hand, bulk water spontaneously freeze at 235 K.\cite{ANGELL:1982aa} Considering all these scattered values, we will use $T_W=235$ K as the Widom temperature to rescale the experimental data and a variation of $\sim 5$ degrees does not affect the validity of the analysis presented in this paper. Therefore, for the purpose of the discussion that follows we assume that the experimental temperatures 235 K and 273.15 K correspond to the temperature of the maximum of the heat capacity and the melting temperature of water, respectively, and a rescaling of the type $\tau=(T-T_W)/(T_m-T_W)$ allows a direct comparison between simulation and experimental results in the supercooled regime, avoiding artifacts due to mistakes in the energy scale of the model systems. However, it should be noted that even if in the rescaled temperature a model reproduces the experimental findings for a given property, an accurate atomistic representation of supercooled water requires also the correct capture, in absolute temperature, of the upper and lower limits of the supercooled regime. The use of the rescaling variable $\tau$ represents a novel way to compare the predictions of different water models that helps to reveal their merits and shortcomings.

In this paper we investigate the water supercooled regime using molecular dynamics simulation and four common water models. Using the rescaling variable $\tau$ we compare the temperature dependence of the heat capacity, diffusion coefficient and hydrogen bonds relaxation time. The analysis of these results in terms of the rescaled variable $\tau$ reveals that the models having explicit lone pairs provide a more credible representation of the supercooled regime. We also use the thermodynamic arguments recently developed by Baumg\"artel and Zimmermann\cite{Baumgartel:2011aa} that relate the difference in enthalpy between water and ice to estimate the critical nucleus for crystallization in the supercooled regime. Finally, we speculate on the reason why spontaneous nucleation is so elusive to atomistic simulations.\cite{Kesselring:2012aa,Shevchuk:2012aa}

\section{Computational Details}
\label{model}

We have performed molecular dynamics simulations using the Gromacs simulation package, v.4.5.5.\cite{Van-der-Spoel:2005aa,Hess:2008aa} Water was described using four widely used  models:  TIP4P-Ew,\cite{Horn:2004aa} TIP4P-2005,\cite{Abascal:2005aa}, TIP5P-Ew\cite{Rick:2004aa} and Six-Sites.\cite{Nada:2003aa} The first two models are four site models, the third contains five sites and the fourth has six interacting sites. The TIP5P-Ew and Six-Sites have a molecular architecture that includes specific sites resembling the water lone pairs. All simulations were done using a cubic simulation box containing 512 molecules and periodic boundary conditions. The temperature of the system was controlled using a Nos\'{e}-Hoover thermostat, with time constant of 0.5 ps. The pressure of the system was controlled by a Parrinello-Rahman barostat, with time constant of 0.5 ps and a compressibility of $4.5 \times 10^{-5}$ bar$^{-1}$. For all models, except for the Six-Sites, we included long-range electrostatic corrections using the PME approach. The leapfrog algorithm was used for the integration of the dynamics equations, with a time-step of 0.001 ps. A spherical cut-off at $r$=0.9 nm was imposed for the Lennard-Jones interactions and short-range electrostatic interactions.

\begin{figure}[!t]
\centering
\includegraphics*[scale=0.55]{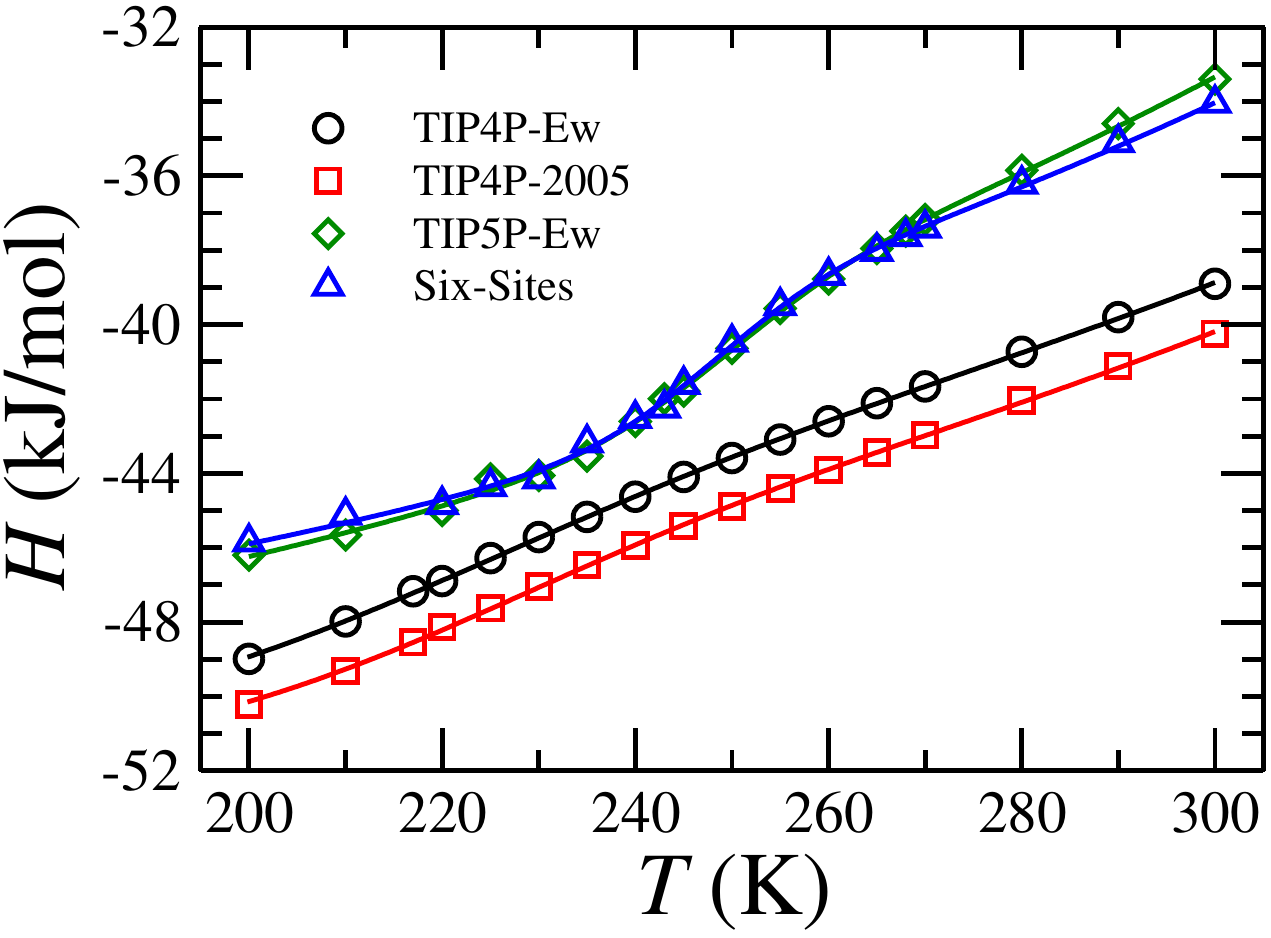}
\caption{Enthalpy as a function of temperature using different water models for supercooled water. The symbols are the results of the simulations, and the lines are fits. The different colors correspond to different models, as displayed in the figure.}
\label{h}
\end{figure}

We simulated liquid water and hexagonal ice in a wide range of temperatures. The liquid high temperature simulations were performed first. The final configuration of each simulation was used as initial configuration for a simulation at the immediate lower temperature. In this way, we achieved a proper equilibration for the coldest systems while at the same time we collect information at the intermediate temperatures. The simulation times range from 20 ns for $T$=300 K to 100 ns for $T$=200 K. The simulations of ice were performed following the same scheme, using a system of 768 molecules as in previous works.\cite{Carignano:2005aa,Carignano:2006aa,Carignano:2007ab} In this case, short 2 ns runs are sufficient to calculate accurate averages of different properties.

\section{Results}
\label{results}

\begin{table*}[!t]
\caption{Fitting parameters for $H$ vs. $T$ using Eq. (\ref{ecuafit}) for water, and a quadratic function for ice.}
\label{tabla-fit}
\begin{ruledtabular}
\begin{tabular}{l|lllll|lll}
Model             &  $A_0$    & $A_1$      & $A_2$     & $A_3$       &    $A_4$ & $B_0$ & $B_1$ & $10^5 B_2$    \\ \hline
TIP4P-Ew      & 2.01037  &  220.736  & 37.9272  &  3.647E-7  &  -50.7219  & -53.281 & - 0.023613 & 17.414\\ 
TIP4P-2005  & 1.66749  &  224.669  & 32.0213  &  3.723E-7  &  -51.909  & -59.973 & 0.023069 & 7.3462\\
TIP5P-Ew      & 1.62349  &  249.491  & 16.1677  &  5.079E-7  &  -48.6777  & -44.344 & - 0.069571 & 25.636 \\
Six-Sites        & 1.67823  &  248.228  & 14.1442  &  4.475E-7  &  -47.7965  & -50.063 & - 0.019388 & 15.476\\
\end{tabular}
\end{ruledtabular}
\end{table*}

\begin{figure}[!b]
\centering
\includegraphics*[scale=0.55]{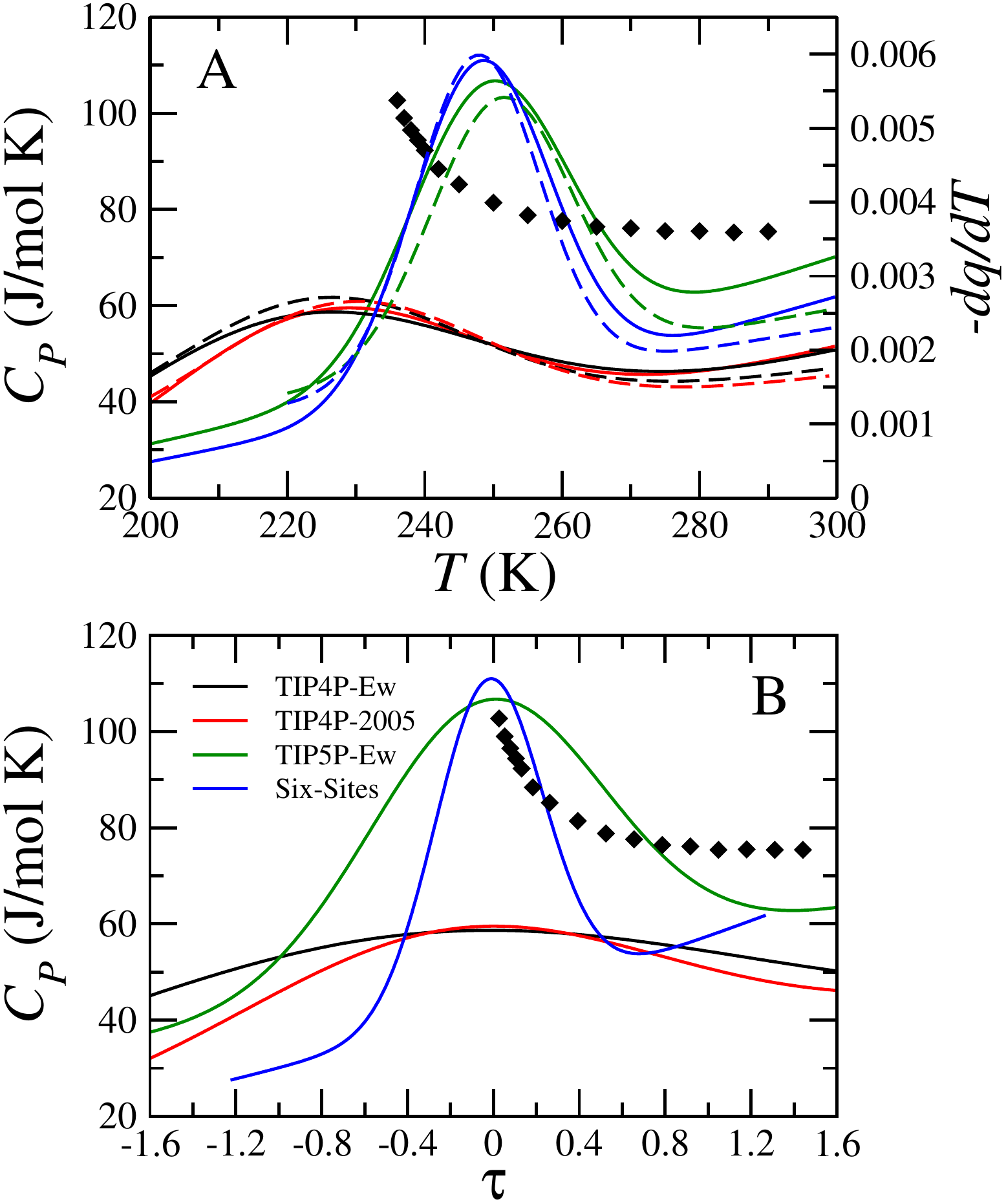}
\caption{A) Isobaric heat capacity (solid lines) and negative of the temperature derivative of the tetrahedral order parameter (dashed lines) as a function of temperature for different water models. The symbols correspond to experimental data from Angell et al.\cite{ANGELL:1982aa} B) Heat capacity as a function of the scaled variable $\tau$.}
\label{cp}
\end{figure}

The calculation of the isobaric heat capacity $C_P$ from $NPT$ simulations can be done in two different ways: {\em i)} from the fluctuation of the enthalpy or {\em ii)} by numerical differentiation of a fitting function of the enthalpy-temperature data. While both methods provide a similar output, the second is preferred because it reduces the noise of the results and allows for an accurate determination of the temperature correspoding to the maximum in $C_P$. For that, it is important to chose a fitting function that captures the essential features of the enthalpy curve. Several previous publications have used a polynomial fit of the enthalpy data.\cite{Poole:2005aa,Paula-Longinotti:2011aa} Here we propose the following fitting function:
\begin{equation}
g\left ( T \right )= A_0 \, \text{erf} \left ( \frac{T-A_1}{A_2}\right )+A_3 T^3+A_4 \,\,\, ,
\label{ecuafit}
\end{equation}
where the erf function targets the main structure of the curve and $A_1$ provides a quick estimation of the temperature corresponding to the maximum in the heat capacity, $T_{C_P}$, which we use as our measure of $T_W$. In Figure \ref{h} we show the results from our simulations of supercooled water for the enthalpy vs temperature relation using the four different models and the corresponding fit using Eq. (\ref{ecuafit}) that in all cases has a correlation coefficient larger than 0.999. The corresponding enthalpy results for hexagonal ice (not shown) display an almost linear relation with the temperature and are very well fitted using a quadratic polynomial $B_0 + B_1 T + B_2 T^2$. The ice enthalpy will be used below to calculate the critical nucleus size for crystallization. The fits for the case of supercooled water were performed using all the data shown in Figure \ref{h}, but for the ice the fits were restricted to temperatures between $T_{C_P}$ and $T_m$. The parameters of the fits are summarized in Table \ref{tabla-fit}. 

\begin{figure}[!t]
\centering
\includegraphics*[scale=0.55]{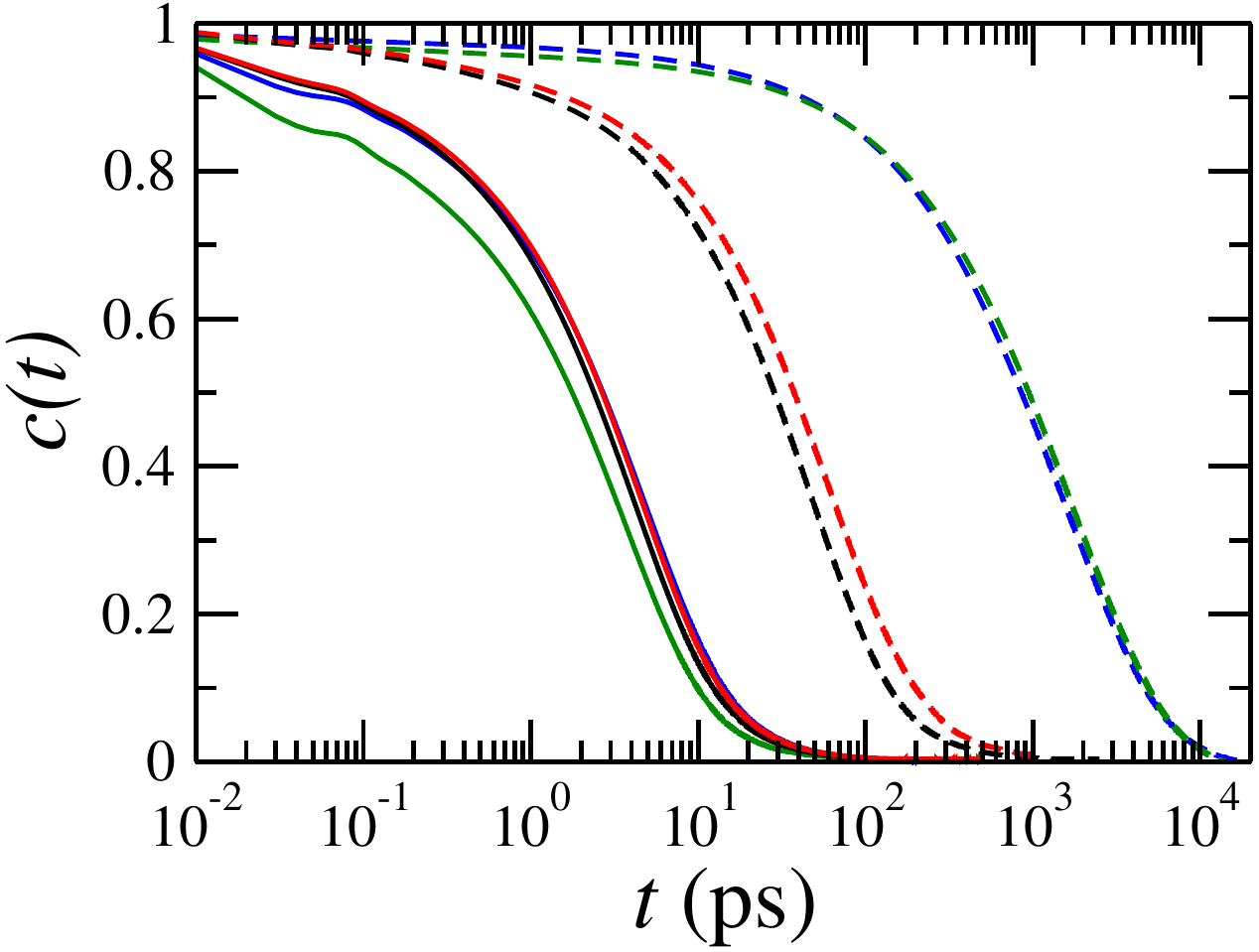}
\caption{History independent hydrogen bond correlation function calculated at 300 K (solid lines) and 240 K (dashed lines). The functions $c(t)$ have been corrected to account for the finite size effects, following \cite{Spoel:2006aa}. TIP4P-Ew (black), TIP4P-2005 (red), TIP5P-Ew (green) and Six-Sites (blue).}
\label{cdet}
\end{figure}

In Figure \ref{cp}A we display the results for $C_P$ for the different models, along with the experimental values from Angell et al.\cite{ANGELL:1982aa}. All the models yield a maximum in the heat capacity. The two four sites models have a similar behavior and show a small difference in the temperature corresponding to the maximum. The other two models have the maximum $C_P$ at higher temperatures. The experimental data show the apparent divergency of $C_P$ at an intermediate temperature between the two groups of models. It is interesting to note that the models with explicit lone pair sites (TIP5P-Ew and Six-Sites) display a sharper peak than the four sites models, presumable due to their tendency to form tetrahedral structures enhanced by the particular molecular architecture. The melting temperaure for all the models and the temperature corresponding to the maximum in $C_P$ calculated by differentiation of Eq. (\ref{ecuafit}) are summarized in Table \ref{tempes}. In Figure \ref{cp}B we display $C_P$ as a function of the scaled temperature $\tau$. This representation shows how the four site models have a very weak temperature dependence, with the peak spreading over the complete supercooled region.

\begin{table}[b]
\caption{Melting temperature ($T_m$) and temperature of the maximum $C_p$ ($T_{C_p}$) for the different models.}
\label{tempes}
\begin{ruledtabular}
\begin{tabular}{lllll}
                      &  TIP4P-Ew & TIP4P-2005 &   TIP5P-Ew  & Six-Sites     \\ \hline 
$T_m$         &    244         &  252             &    271           &    289    \\   
$T_{C_p}$  &    227         &  229             &    250           &    249    \\ 
\end{tabular}
\end{ruledtabular}
\end{table}

The structure of the liquid can be characterized by the tetrahedral order parameter:\cite{Errington:2001aa}
\begin{equation}
q(k)=1-\frac{3}{8} \sum_{i=1}^3 \sum_{j=i+1}^4 \left ({\cos{\psi_{ikj}}+\frac{1}{3}} \right)^2  \,\,\,.
\end{equation}
This parameter is calculated for every molecule $k$, in terms of the angle $\psi_{ikj}$ defined by the lines joining the oxygen atom of the central molecule $k$, and those of its four nearest neighbors. $q$ takes values in the range $-3 \leq q \leq 1$. For perfect tetrahedral order, $q(k) = 1$; and for random molecular order, $\langle q(k) \rangle = 0$. The tetrahedral order parameter monotonically increases as the temperature decreases.\cite{Guillermo-Pereyra:2012aa} Moreover, the curve $-q$ vs $T$ (not shown) is remarkably similar to $H$ vs $T$ and consequently the derivative  $-dq/dT$ closely follows the heat capacity, as shown in Figure \ref{cp}A. The striking similarity of these two curves, $C_P(T)$ and $-dq/dT$, indicates a very strong correlation between the thermodynamics of the system and its structure, in this case characterized by the local tetrahedral order. The enthalpy of the system is dominated by the potential energy term, which in turn is dominated by the local tetrahedral order.\cite{Guillermo-Pereyra:2012aa} The maxima in $C_P$ and in $-dq/dT$ reflect the change in curvature of the enthalpy and order parameters curves, both occurring at approximately the same temperature. More tetrahedral arrangements occur for $T < T_{C_P}$ favoring the LDL phase, and lower tetrahedral order characterizes the other side of the peak favoring the HDL phase.

\begin{figure}[!t]
\centering
\includegraphics*[scale=0.55]{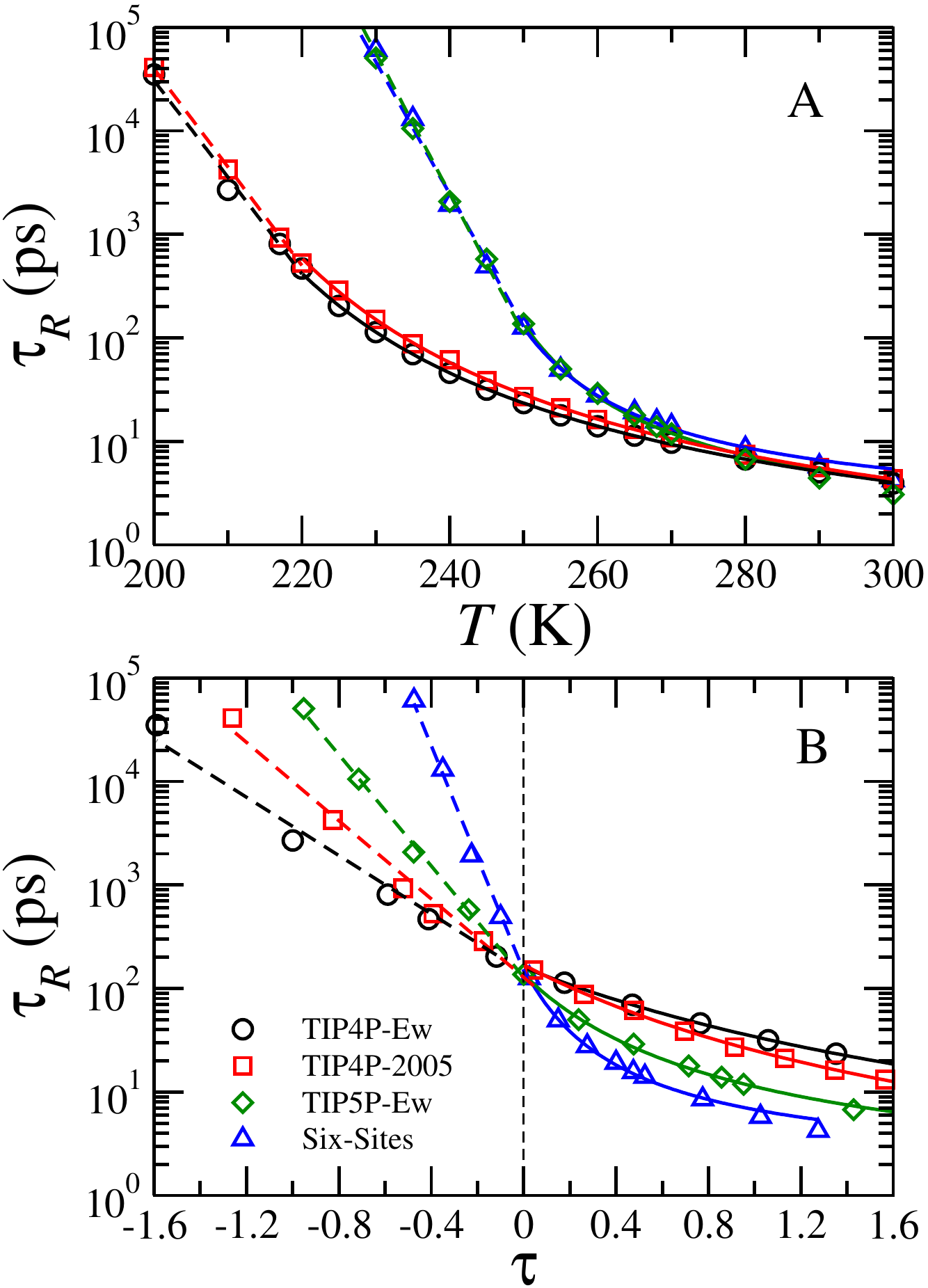}
\caption{Hydrogen bonds relaxation times for the different models  as a function of A) temperature and B) scaled temperature $\tau$. The solid lines are fitting to the simulation data using the VTF function and dashed lines correspond to an Arrhenius fit.}
\label{taus}
\end{figure}

We now turn our attention to dynamical properties, and start the analysis by considering the kinetics of the hydrogen bonds. We use a geometrical criterium to define hydrogen bonds: two molecules are bonded if their O-O distance is smaller than the distance corresponding to the first minimum of the radial distribution function, and if the O $\cdots$ H-O angle is smaller than 30$^\circ$. The history-independent hydrogen bond correlation function is defined by \cite{Luzar:1996ab,Luzar:1996aa}
\begin{equation}
c(t) = \frac{ \langle {h_i}_j(0){h_i}_j( t ) \rangle }{ \langle h \rangle}  \,\,\, ,
\label{ecuahb}
\end{equation}
where ${h_i}_j(t)=1$ ($=0$) when the molecules $i$ and $j$ at a time $t$ form (do not form) a hydrogen bond, $\langle \rangle$ denotes the average over all the pairs $ij$, and $ \langle h \rangle$ denotes the average of the operator $h$ over all the pairs and all the times. The function $c(t)$ provides information about the lifetime of the hydrogen bonds, since its value represents the proportion of hydrogen bonds that remain as such after a time $t$. Examples of the hydrogen bonds correlation function is displayed  in Figure \ref{cdet} for the four models and two different temperatures. While for 300 K all the models have a similar relaxation time, for 240 K the TIP4P models relax much faster than the other two models.\cite{Guillermo-Pereyra:2012aa} The relaxation times $\tau_R$, defined by the condition $c(\tau_R)=e^{-1}$, are displayed in Figure \ref{taus}A for all the models and as a function of the temperature. The figure also includes the lines obtained by fitting the data with the empirical  Vogel-Tammann-Fulcher (VTF) function $\tau_R(T)=A_0 \exp(A_1/(T-T_0))$ for $T > T_{C_P}$ and an Arrhenius expression $\tau_R(T)=M_0 \exp(-E_a/RT)$ for $T < T_{C_P}$. These two different functional forms are often used to analyze the fragile to strong transition that also occurs at $T_{C_P}$, as it is extensively explained in numerous publications.\cite{Mallamace:2006aa} In Figure \ref{taus}B we show the same relaxation times now as a function of the scaled variable $\tau$. The plots illustrate the weak temperature dependence of the four sites models but more remarkably, all the models have approximately the same relaxation time ($\tau_R \simeq 200$ ps) at the temperature corresponding to the maximum of the heat capacity. The agreement of the relaxation times at $\tau=0$ will be discussed below in connection with the diffusion coefficients of the four models studied in this work. The parameters for the fits of Figure \ref{taus} are summarized in Table \ref{taufit}.

\begin{table}[b]
\caption{Parameters of the best fits of the hydrogen bonds relaxation times using the VTF  and Arrhenius equations, as described in the text.}
\label{taufit}
\begin{ruledtabular}
\begin{tabular}{lllll}
                   &  TIP4P-Ew & TIP4P-2005 &   TIP5P-Ew  & Six-Sites     \\ \hline 
$T_0$       &    174.22       &  172.52          &    229.82           &    231.48    \\   
$A_0$       &    0.28   &    0.23  &       1.02   &   1.68   \\ 
$A_1$       &   333.79     &   371.74       &   98.61         &  79.78 \\
\hline
$M_0$  &    $8.917 \times 10^{-17}$    &  $5.546 \times 10^{-17}$      &    $4.714 \times 10^{-28}$     &    $6.114 \times 10^{-30}$    \\   
$E_a  $       &   78.68    &   79.91    &     140.99     &    149.69 \\ 
\end{tabular}
\end{ruledtabular}
\end{table}

\begin{figure}[!t]
\centering
\includegraphics*[scale=0.3]{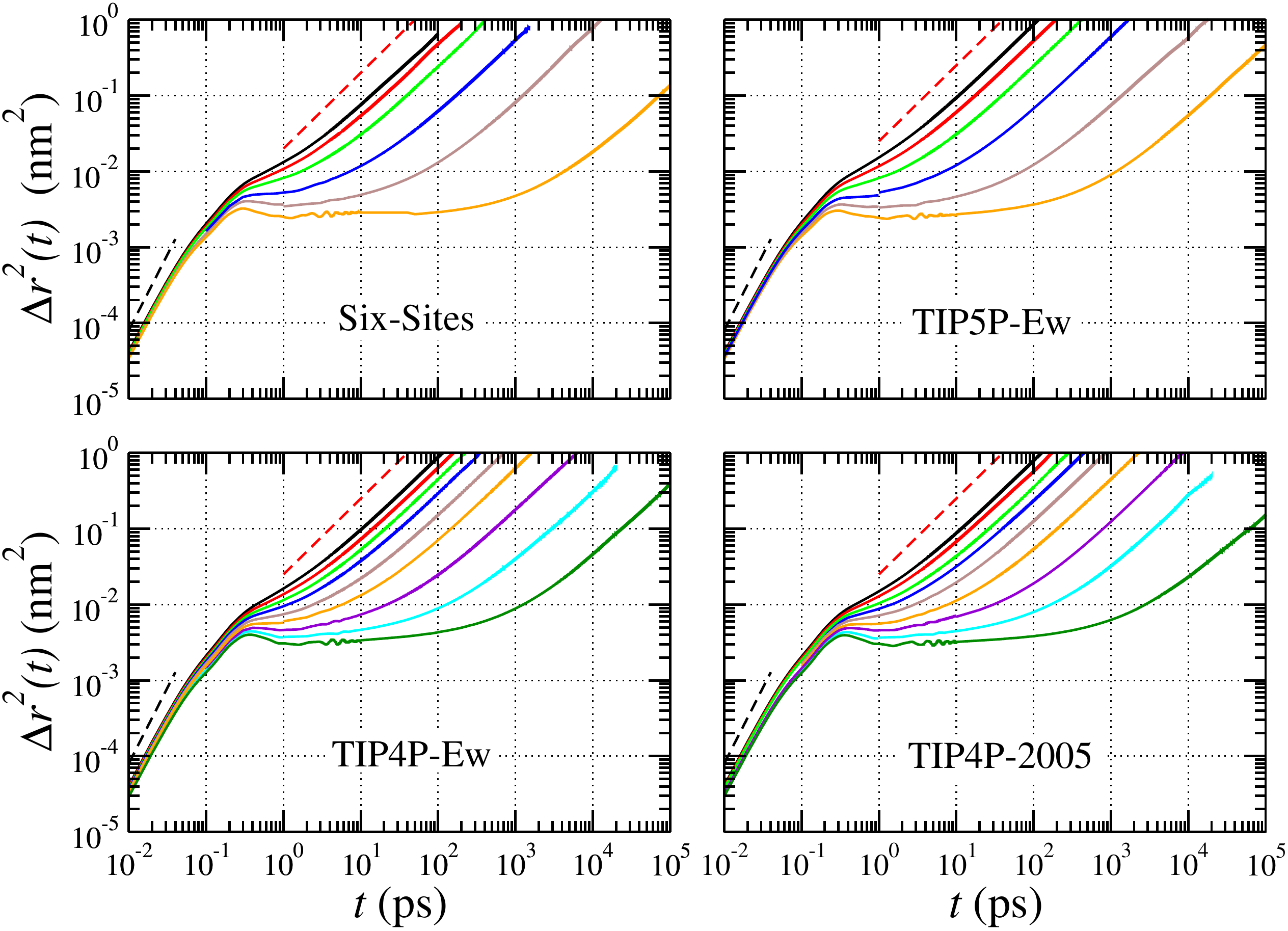}
\caption{Mean square displacement as a function of time, for all the models. Different colors correspond to different temperatures as follows: 280 K (black), 270 K (red), 260 K (green), 250 K (blue), 240 K (brown), 230 K (orange), 220 K (violet), 210 K (cyan) and 200 K (dark green). The black and red dashed lines represent the ideal ballistic and diffusive limits, respectively.}
\label{cage}
\end{figure}

The calculated relaxation times indicate an important slow down in the kinetics of the system as the temperature is decreased below the temperature of the maximum of the heat capacity, $T_{C_p}$. To further analyze this effect, we consider the mean squared displacement $\Delta r^2(t)=\langle ( r(t)-r(0) )^2 \rangle $, as a function of the time interval, for the different temperatures. For a diffusive process, $\Delta r^2(t)=6 D_0 t$, with $D_0$ being the diffusion constant. This is the observed behavior for liquid water, for $t$ sufficiently large. For very short times, $\Delta r^2(t) \propto t^2$ reflects the ballistic motion of the molecules between collisions. For low temperatures, our molecular dynamics simulations reveal a plateau regime between the ballistic and the diffusive limits, corresponding to the so-called caging effect.\cite{Sciortino:1996aa} On Figure \ref{cage} we show the results for all the models, at several temperatures. Notice that in the double logarithmic plot, the ballistic and diffusive limits are characterized by a slope of 2 and 1, respectively. In all the cases, the caging effect is observed for $T$ smaller than the melting temperature. However, for $T$ smaller than the temperature of the maximum of $C_P$ there is a dramatic increase in the length of the plateau region. These results are showing essentially the same phenomenon indicated by the temperature dependence of the relaxation times $\tau_R$, as the system exhibit an important slowdown for temperatures smaller the $T_{C_P}$

\begin{figure}[!t]
\centering
\includegraphics*[scale=0.55]{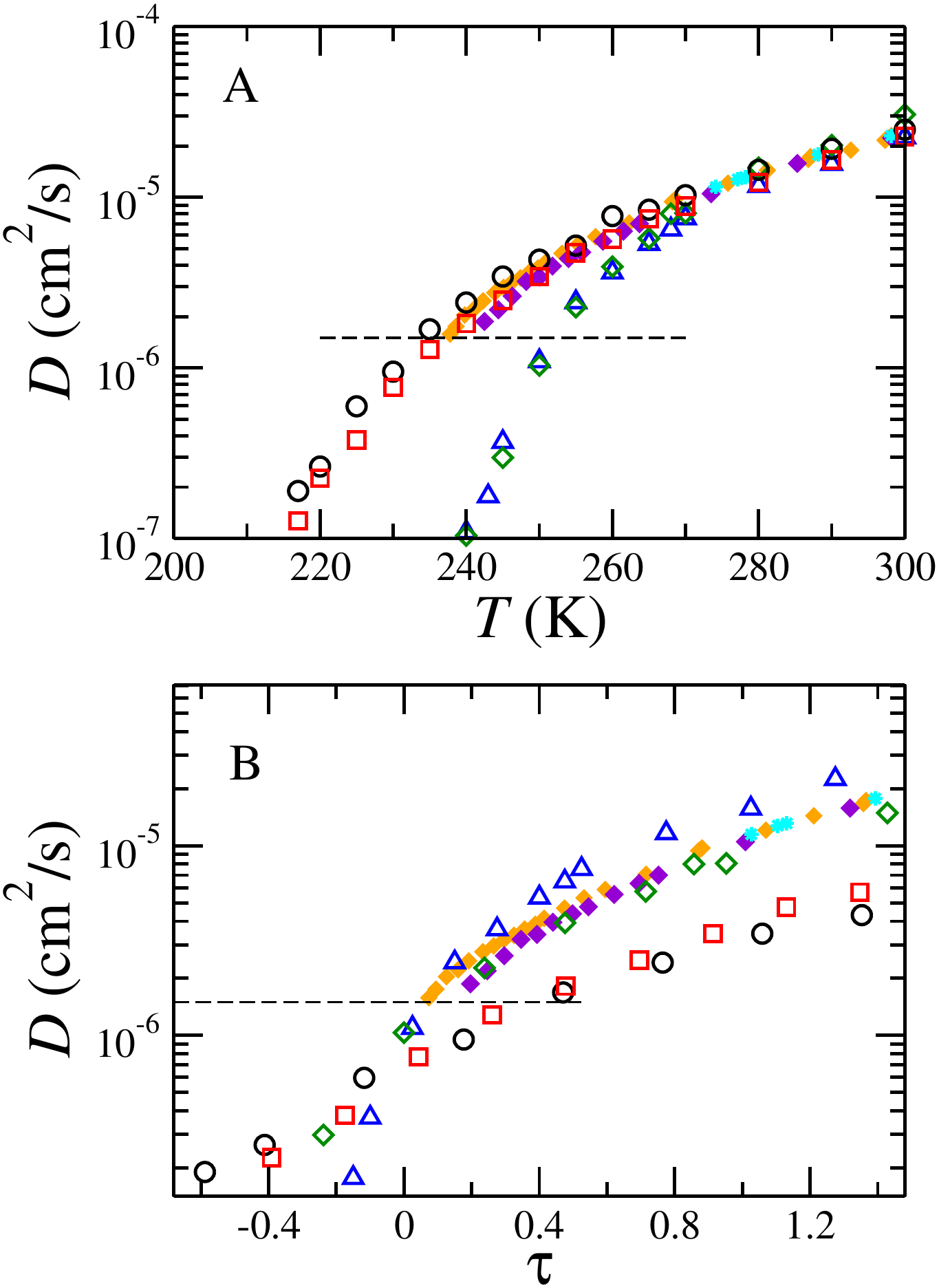}
\caption{Self diffusion coefficient as function of A) temperature and B) rescaled temperature $\tau$ for differents water models (open symbols) and experimental data (full symbols) from Price et al.\cite{Price:1999aa} (orange), Gillen et al. \cite{GILLEN:1972ac} (violet) and Mills\cite{MILLS:1973aa} (cyan). The dashed horizontal line indicates the lowest experimental diffusion values. TIP4P-Ew (black circles), TIP4P-2005 (red squares), TIP5P-Ew (green diamonds) and Six-Sites (blue triangles).}
\label{difusion}
\end{figure}

In Figure \ref{difusion}A we show the diffusion coefficients calculated from diffusive branch of the mean square displacement, for all the models and temperatures simulated. The figure also includes experimental values of the diffusion coefficient in bulk water obtained by different authors. The reduction (vanishing) of $D_0$ with decreasing temperature is another manifestation of the dramatic increase of the relaxation times as the temperature decreases below $T_{C_P}$. At first sight, it is striking the similarity between the prediction of the TIP4P models and the experimental values. However, the value of such agreement is diminished by the fact that the melting temperature of the TIP4P models is lower than the experimental one, see Table \ref{tempes}. Therefore, for most of the experimental range, the TIP4P models are in a  stable liquid phase, and do not represent metastable supercooled water.  Another fact immediately apparent from Figure \ref{difusion}A is that the simulations yield diffusion coefficients much smaller than the smallest experimental value in the bulk, represented by the dashed horizontal line. Interestingly, the intersection of this line with the simulation results occur at a temperature very close to $T_{C_P}$ for the corresponding model as can be clearly seen from Figure \ref{difusion}B that displays the diffusion coefficients as a function of $\tau$. It could be argued that the metastable liquid has a minimum mobility below which it spontaneously transforms to the stable crystal phase, and this transformation occurs when the hydrogen bond network has relaxation times longer than 200 ps, as suggested by Figure \ref{taus}B.

Supercooled water eventually crystalize to hexagonal ice. This spontaneous crystallization is a complex phenomenon that starts from the random assembly of embryo nuclei, some of which will succeed and grow to macroscopic size and some will fail and melt depending whether they are larger or smaller than a critical nucleus size, respectively. The number of successful nucleus formations per unit time and volume is the homogeneous nucleation rate ($J$) and its value can be experimentally determined.\cite{HUANG:1995aa,Kramer:1999aa,Stockel:2005aa,Kabath:2006aa,Duft:2004aa,Baumgartel:2011aa,pruppacherklett,Pruppacher:1995aa,Jeffery:1997aa,HAGEN:1981aa} It is observed that $J$ is extremely small for temperatures above 244 K, but its value increases dramatically for lower temperatures. The strong temperature dependence explains the discrepancies of up to one order of magnitude in the experimental determination of $J$ among different authors. For example, at $T=237$ K different authors report very different values for $J$. Huang et al.\cite{HUANG:1995aa}, Kr\"amer et al.\cite{Kramer:1999aa}, St\"ockel et al.\cite{Stockel:2005aa}, Kabath et al.\cite{Kabath:2006aa} and Baumg\"artel and Zimmermann\cite{Baumgartel:2011aa}, have reported $J$ = 1, 30, 4, 8.13 and 2.1 ($\times 10^5 $cm$^{-3}$s$^{-1}$), respectively.

From the simulation point of view, direct quantification of size of the critical nucleus seems to be an impossible task due the time scale of the process. Using an indirect approach, Pereyra et al.\cite{Pereyra:2009aa,Pereyra:2011aa} have studied the stability of ice nanocolumns in vacuum and liquid water and they have determined the minimal radius of the growing nanocolumns. Using the Gibbs-Thomson equations, the critical nucleus size for 3D particles was deduced from the 2D simulation results. In this paper we exploit the idea presented by Baumg\"artel and Zimmermann\cite{Baumgartel:2011aa} who  measured $J$ in a small range of temperatures (30 measurements within 1 degree) around 237 K and performed and Arrhenius analysis of the $J$ vs $T$ relation, finding an activation energy $E_a=-2.68 \times 10^{-18}$ J. Next, they propose the hypothesis that the activation energy is equal to the change of enthalpy between the liquid and crystalline phases of the molecules involved in the critical nucleus. Namely
\begin{equation}
E_a=N \Delta H_c
\label{ea}
\end{equation}
where $N$ is the critical nucleus size and $\Delta H_c$ is the enthalpy of crystallization per one water molecule. For the final numerical estimation of $N$, Baumg\"artel and Zimmermann \cite{Baumgartel:2011aa} used $\Delta H_c$ from a fourth order polynomial expression given by Pruppacher and Klett{\cite{pruppacherklett}. In our approach, we start from the known experimental values for the nucleation rate in the whole supercooled regime\cite{pruppacherklett,Pruppacher:1995aa}, which are plotted in Figure \ref{jota}A. The data is clearly non-Arrhenius in the whole temperature range, and in particular for $T=237$ K. However, the low temperaure regime follows an Arrhenius curve with activation energy of $-2.56 \times 10^{-18}$ J. The higher temperature branch is well capture with a VTF function $J=\exp(A+B/(T-T_0))$, with $A=106.844$, $B=2177.49$ and $T_0=261.074$. Following Ediger et al.\cite{Ediger:1996aa} we define an {\em apparent} activation energy by
\begin{equation}
E_{a}=-k_B \frac{d \ln J}{d  (1/T)}=-k_B B \left ( 1-\frac{T_0}{T}\right )^{-2}
\label{ediger}
\end{equation}
that will later be related to $\Delta H_c$ in order to calculate the critical nucleus size for ice formation. In Figure \ref{jota}B we show the Arrhenius and VTF (apparent) activation energies as a function of temperature. With this result for $E_a$, combined with $\Delta H_c$ calculated from simulations and using Eq. (\ref{ea}) we obtain an estimation for the critical nucleus size as a function of temperature, displayed on Figure \ref{nucleus} as a function of the scaled variable $\tau$. Note that the critical nucleus size predicted by the four site models is significantly larger than the corresponding to the TIP5P-Ew and Six-Sites models. Again in this case, the models having explicit lone pairs have a better agreement with the available experimental value.

\begin{figure}[!t]
\centering
\includegraphics*[scale=0.55]{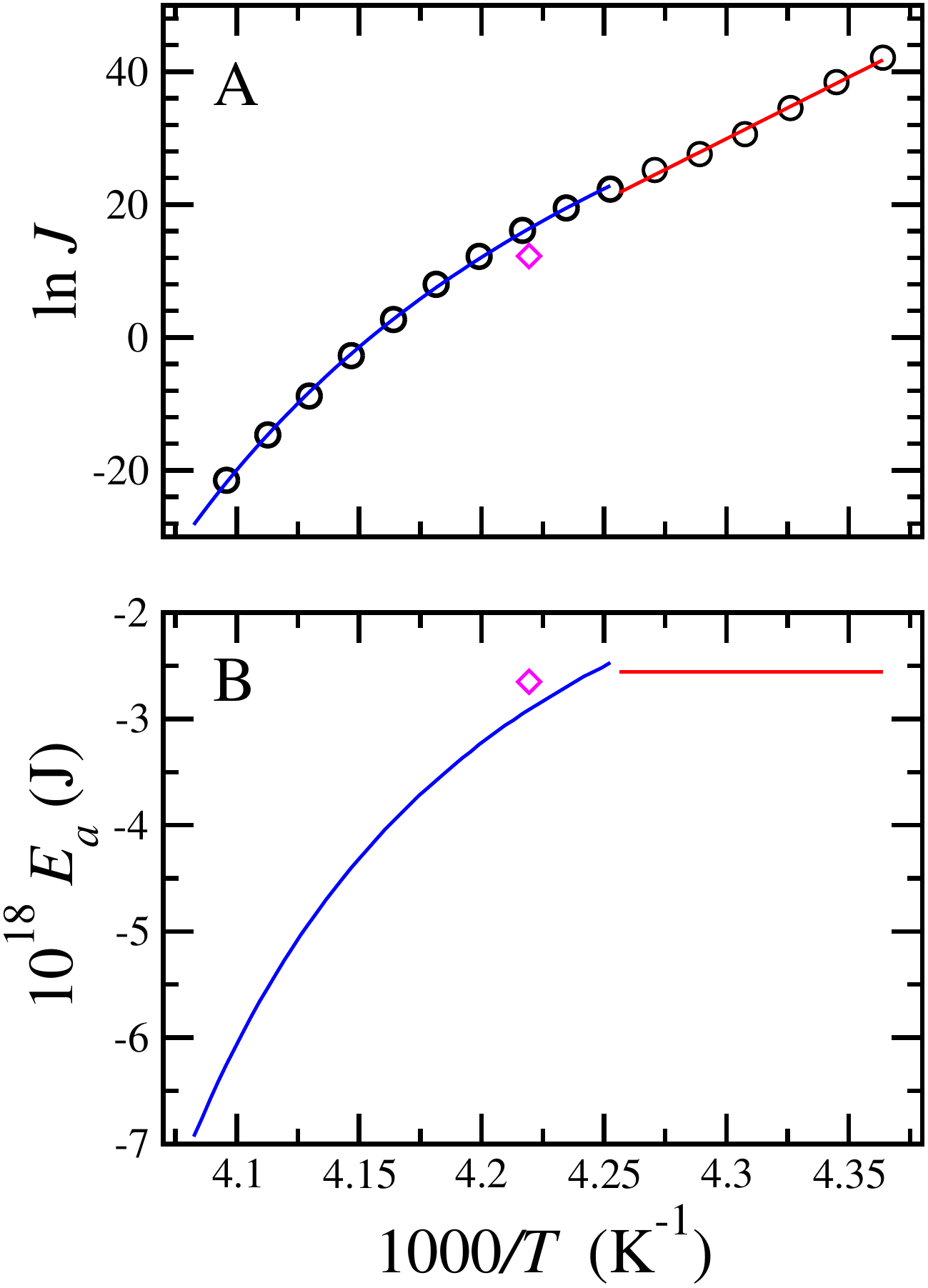}
\caption{A) Experimental homogeneous nucleation rate from Pruppacher et al.\cite{pruppacherklett,Pruppacher:1995aa} (circles) and Baumg\"artel and Zimmermann\cite{Baumgartel:2011aa} (diamond). The lines represent VTF (blue) and Arrhenius plots (red). B) Apparent activation energy calculated calculated using Eq. (\ref{ediger}) (blue), and Arrhenius equation (red).}
\label{jota}
\end{figure}

\section{Discussion}
\label{final}

The comprehensive analysis of all the results presented in this paper reveals that the TIP5P-Ew model provides a better representation of supercooled water than the other three models. The agreement of the diffusion coefficient calculated with the four sites models and the experimental values in terms of absolute temperature breaks down when the temperature scale is corrected by the temperatures that define both ends of the supercooled regime. The models having explicit lone pairs behave in a similar way, but the TIP5P-Ew has a better agreement with experiments in terms of the absolute temperature and therefore is preferred over the Six-Sites model.  

\begin{figure}[!t]
\centering
\includegraphics*[scale=0.55]{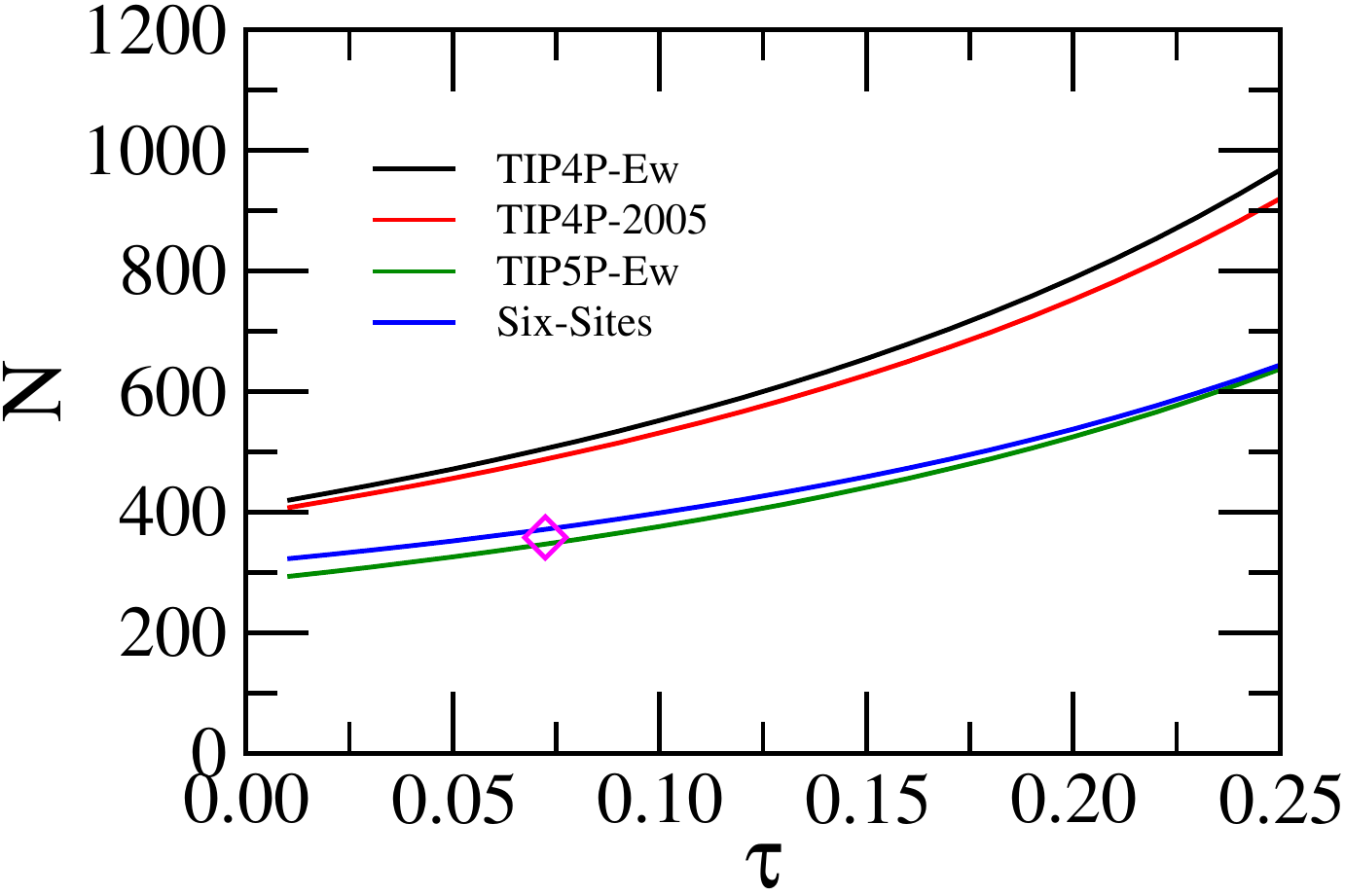}
\caption{Critical nucleus size for ice growth as a function of the scaled temperature $\tau$. The lines represent the prediction of the different models, and the diamond corresponds to the measurement of Baumg\"artel and Zimmermann.\cite{Baumgartel:2011aa}}
\label{nucleus}
\end{figure}

The question that remains is why the simulated systems do not spontaneously crystallized after many trials of approximately 1 $\mu s$ of simulation time.\cite{Kesselring:2012aa,Shevchuk:2012aa} It is clear that as the temperature decreases the system monotonically increases its tetrahedral order while at the same time dramatically increasing the hydrogen bonds relaxation times and decreasing the molecular mobility. The maximum in the heat capacity, reflecting the largest enthalpy fluctuations, occurs at a temperature where the tetrahedral fluctuations are also maximal. For temperatures smaller than $T_{C_p}$, all fluctuations start to die out. Recent simulation results by Moore and Molinero based on the coarse grained mW model do crystallize in typical simulation times.\cite{Moore:2010aa,Moore:2011aa} For a high cooling rate, the supercooled mW HDL transforms to a LDL state as the temperature decreases. For a low cooling rate, the supercooled mW liquid spontaneously crystallizes. The enthalpy vs. temperature curves clearly show that these two transitions occur at the same temperature, i.e. at the temperature corresponding to the maximum of $C_P$. Although there are alternative explanations,\cite{Shevchuk:2012aa} the reason why the mW model is able to crystallize is, in our opinion, due to its core repulsive potential that much softer than the typical $r^{-12}$ of atomistic models.
Therefore, the mW model is able to squeeze and rotate toward an organized, crystalline structure. One way in which atomistic models could incorporate an effective softer repulsion is to allow for proton transfer between molecules. In fact, recent experimental findings on proton diffusivity and characteristic hopping times support this idea. Presiado et al.\cite{Presiado:2011aa} measured that the proton hopping time in hexagonal ice doped with 1 mM HCl increases from 1 ps at 225 K to 1 ns at 140 K. Then, considering that the kinetics of supercooled water becomes very slow as reveled by the hydrogen bonds relaxation times and diffusion coefficient, the proton hopping mechanism is at two order of magnitude faster and therefore could dominate the overall system dynamics. Therefore it is worth to explore the possibility of proton transfer as an auxiliary mechanism able to unlock highly jammed tetrahedral structures; in other words, providing an effectively softer hard core repulsion between molecules.

\section{acknowledgments}
Marcelo Carignano acknowledges the support from NSF (grant CHE-0957653).

%\pagebreak 

%\bibliographystyle{achemso}
%\bibliography{feb2011}

\end{document}